# Solution of the Dirac equation with non-minimal coupling to noncentral three-vector potential

A. D. Alhaidari

*Physics Department, King Fahd University of Petroleum & Minerals, Dhahran 31261, Saudi Arabia*
e-mail: haidari@mailaps.org

We introduce non-minimal coupling to three-vector potential in the 3+1 dimensional Dirac equation. The potential is noncentral (angular-dependent) such that the Dirac equation separates completely in spherical coordinates. The relativistic energy spectrum and spinor wavefunctions are obtained for the case where the radial component of the vector potential is proportional to $\frac{1}{r}$. The non-minimal coupling presented in this work is a generalization of that which was introduced by Moshinsky and Szczepaniak in the Dirac-Oscillator problem.

## I. INTRODUCTION

The Dirac equation is a relativistically covariant first order linear differential equation in space and time. It describes a spinor particle at relativistic energies below the threshold of pair production. It also embodies the features of quantum mechanics as well as special relativity. However, despite all the work that has been done over the years on this equation, its exact solution has been limited to a very small set of potentials. Since the original work of Dirac in the early part of last century up until 1989 only the relativistic Coulomb problem was solved exactly. In 1989, the relativistic extension of the oscillator problem (Dirac-Oscillator) was formulated and solved by Moshinsky and Szczepaniak [1]. Recently, an effective approach for solving the Dirac equation with spherical symmetry was introduced [2-4]. It started with the realization that the nonrelativistic Coulomb, Oscillator, and S-wave Morse problems belong to the same class of potentials which carries a representation of so(2,1) Lie algebra. Therefore, the fact that the relativistic version of the first two problems (Dirac-Coulomb and Dirac-Oscillator) was solved exactly makes the solution of the third, in principle, feasible. Indeed, the relativistic Dirac-Morse problem was formulated and solved in Ref. 2. The bound state energy spectrum and spinor wavefunctions were obtained. Taking the nonrelativistic limit reproduces the familiar Schrödinger-Morse problem. The same approach was applied successfully in obtaining solutions for the relativistic extension of yet another class of shape invariant potentials. These included the Dirac-Scarf, Dirac-Rosen-Morse I & II, Dirac-Pöschl-Teller, Dirac-Eckart, Dirac-Hulthén, and Dirac-Woods-Saxon potentials [3]. Furthermore, using the same formalism quasi exactly solvable systems at rest mass energies were obtained for a large class of power-law relativistic potentials [4]. $L^2$ series solutions of the Dirac equation for scattering and bound states were obtained in terms of orthogonal polynomials for several interactions [5]. The Dirac-Coulomb problem with a relativistic singular mass distribution (position-dependent effective mass) was also solved [6]. Quite recently, a systematic and intuitive approach for the separation of variables in the Dirac equation with coupling to noncentral electromagnetic potential was introduced [7]. The Dirac-Coulomb problem in the presence of the Aharonov-Bohm effect and magnetic monopole potential was solved. The same formulation was extended to the study of the relativistic Dirac-Oscillator in the presence of noncentral electromagnetic potential [8]. In this article we utilize these findings in obtaining a solution of the Dirac equation in 3+1 dimensions with non-minimal coupling to the three-vector potential $\vec{A}(\vec{r})$ whose components are

$$A_r = \frac{\lambda}{r}, \quad A_\theta = \frac{b\cos\theta - a}{r\sin\theta}, \quad A_\phi = 0, \tag{1.1}$$

where $\lambda$, $a$, and $b$ are real dimensionless parameters. If $b = \lambda$, then this potential satisfies the Coulomb gauge, $\vec{\nabla} \cdot \vec{A} = 0$. The coupling of the Dirac spinor to the vector potential is a generalization of that which was introduced by Moshinsky and Szczepaniak in the solution of the Dirac-Oscillator problem. It should, however, be noted that this type of coupling does not support an interpretation of $\vec{A}(\vec{r})$ as the magnetic vector potential. In fact, one can easily verify that such an interpretation results in zero magnetic field since $\vec{\nabla} \times \vec{A} = 0$.

In the following section, we write the Dirac equation in spherical coordinates with the non-minimal coupling to the vector potential and show that it separates completely into radial and angular components. The angular Dirac equation will be solved in Sec. III, whereas in Sec. IV we solve the radial Dirac equation and calculate the relativistic energy spectrum. Additionally, the four-component spinor wavefunction will be obtained.

## II. COUPLING TO THE VECTOR POTENTIAL AND SEPARATION OF VARIABLES

In the relativistic units, $\hbar = c = 1$, the Dirac equation in 3+1 dimensions for a free spinor reads $\left(i\gamma^\mu \partial_\mu - M\right)\Psi = 0$, where $M$ is the rest mass of the particle and $\Psi$ is the four component spinor wavefunction. $\partial_\mu = \left(\frac{\partial}{\partial t}, \vec{\nabla}\right)$ is the space-time gradient and $\{\gamma^\mu\}_{\mu=0}^3$ are four square matrices with the following standard representation

$$\gamma^0 = \begin{pmatrix} I & 0 \\ 0 & -I \end{pmatrix}, \quad \vec{\gamma} = \begin{pmatrix} 0 & \vec{\sigma} \\ -\vec{\sigma} & 0 \end{pmatrix}, \tag{2.1}$$

where $I$ is the 2×2 unit matrix and $\vec{\sigma}$ are the three 2×2 hermitian Pauli spin matrices:

$$\sigma_1 = \begin{pmatrix} 0 & 1 \\ 1 & 0 \end{pmatrix}, \quad \sigma_2 = \begin{pmatrix} 0 & -i \\ i & 0 \end{pmatrix}, \quad \sigma_3 = \begin{pmatrix} 1 & 0 \\ 0 & -1 \end{pmatrix}. \tag{2.2}$$

Now, we introduce coupling to the three-vector potential $\vec{A}(\vec{r})$ by the non-minimal substitution $\vec{\nabla} \rightarrow \vec{\nabla} + \vec{A}\gamma^0$ (i.e., $\vec{P} \rightarrow \vec{P} - i\vec{A}\gamma^0$, where $\vec{P}$ is the linear three-momentum operator $-i\vec{\nabla}$). This coupling maps the free Dirac equation above into

$$i\frac{\partial}{\partial t}\Psi = \left(-i\gamma^0\vec{\gamma}\cdot\vec{\nabla} + i\vec{\gamma}\cdot\vec{A} + \gamma^0 M\right)\Psi \tag{2.3}$$

For time-independent potential, the spinor wavefunction could be written as $\Psi(t,\vec{r}) = e^{-i\varepsilon t}\psi(\vec{r})$ and the Dirac equation becomes the eigenvalue wave equation $(\mathcal{H} - \varepsilon)\psi = 0$, where $\varepsilon$ is the relativistic energy and $\mathcal{H}$ is the 4×4 Hamiltonian matrix operator

$$\mathcal{H} = \begin{pmatrix} M & -i\vec{\sigma}\cdot\vec{\nabla} + i\vec{\sigma}\cdot\vec{A} \\ -i\vec{\sigma}\cdot\vec{\nabla} - i\vec{\sigma}\cdot\vec{A} & -M \end{pmatrix} \tag{2.4}$$

The special case where $\vec{A} = \vec{r}$ corresponds to the Dirac-Oscillator problem [1]. We write the spinor wave function as $\psi(\vec{r}) = \frac{1}{r\sqrt{\sin\theta}} \begin{pmatrix} if_+(\vec{r}) \\ f_-(\vec{r}) \end{pmatrix}$ and take the components of the vector potential as follows

$$A_r(\vec{r}) = V(r), \quad A_\theta(\vec{r}) = \frac{1}{r} W(\theta), \quad A_\phi(\vec{r}) = \frac{1}{r\sin\theta} Q(\phi) \qquad (2.5)$$

Consequently, the action of the Dirac Hamiltonian (2.4) on the four-component spinor $\begin{pmatrix} f_+ \\ f_- \end{pmatrix}$ is represented by the 4×4 matrix operator

$$\mathcal{H} = \mathcal{H}_0 + \vec{\sigma} \cdot \hat{r} \mathcal{H}_r + \frac{\vec{\sigma} \cdot \hat{\theta}}{r} \mathcal{H}_\theta + \frac{\vec{\sigma} \cdot \hat{\phi}}{r\sin\theta} \mathcal{H}_\phi, \qquad (2.6)$$

where $(\hat{r}, \hat{\theta}, \hat{\phi})$ are the unit vectors in spherical polar coordinates and

$$\mathcal{H}_0 = \begin{pmatrix} M & 0 \\ 0 & -M \end{pmatrix}, \quad \mathcal{H}_\phi = \begin{pmatrix} 0 & -\partial_\phi + Q(\phi) \\ \partial_\phi + Q(\phi) & 0 \end{pmatrix} \qquad (2.7a)$$

$$\mathcal{H}_r = \begin{pmatrix} 0 & -\partial_r + \frac{1}{r} + V(r) \\ \partial_r - \frac{1}{r} + V(r) & 0 \end{pmatrix}, \qquad (2.7b)$$

$$\mathcal{H}_\theta = \begin{pmatrix} 0 & -\partial_\theta + \frac{\cos\theta}{2\sin\theta} + W(\theta) \\ \partial_\theta - \frac{\cos\theta}{2\sin\theta} + W(\theta) & 0 \end{pmatrix}. \qquad (2.7c)$$

Square integrability and the boundary conditions require that the spinor components satisfy: $f_\pm(r)|_{r=0}$ and $f_\pm(\theta)|_{\theta=0,\,\theta=\pi}$ are finite, $f_\pm(r)|_{r\to\infty} = 0$ and $f_\pm(\phi) = f_\pm(\phi + 2\pi)$. To simplify the solution, we employ a 2×2 unitary transformation matrix $\Lambda(\hat{r})$ that maps the spherical projection of the Pauli matrices ($\vec{\sigma}\cdot\hat{\theta}, \vec{\sigma}\cdot\hat{\phi}, \vec{\sigma}\cdot\hat{r}$) into their canonical Cartesian representation ($\sigma_1, \sigma_2, \sigma_3$), respectively. That is,

$$\Lambda^{-1} \vec{\sigma} \cdot \hat{\theta} \Lambda = \sigma_1, \quad \Lambda^{-1} \vec{\sigma} \cdot \hat{\phi} \Lambda = \sigma_2, \quad \Lambda^{-1} \vec{\sigma} \cdot \hat{r} \Lambda = \sigma_3. \qquad (2.8)$$

Other permutations of the $\sigma_i$'s on the right differ only by a unitary transformation. The following unitary matrix achieves the goal

$$\Lambda(\hat{r}) = e^{-\frac{i}{2}\sigma_3 \phi} e^{-\frac{i}{2}\sigma_2 \theta}. \qquad (2.9)$$

Thus, the transformed wavefunction, which we write as $\chi = \begin{pmatrix} g_+ \\ g_- \end{pmatrix}$, has the two-component spinors $g_\pm = \Lambda^{-1} f_\pm$, whereas the matrix operator (2.6) gets mapped into the following hermitian representation of the Dirac Hamiltonian

$$H = H_0 + \sigma_3 H_r + \frac{\sigma_1}{r} H_\theta + \frac{\sigma_2}{r\sin\theta} H_\phi, \qquad (2.10)$$

where $H_0 = \mathcal{H}_0$, $H_\phi = \mathcal{H}_\phi$ and

$$H_r = \begin{pmatrix} 0 & -\partial_r + V(r) \\ \partial_r + V(r) & 0 \end{pmatrix}, \quad H_\theta = \begin{pmatrix} 0 & -\partial_\theta + W(\theta) \\ \partial_\theta + W(\theta) & 0 \end{pmatrix}. \qquad (2.11)$$

Therefore, we obtain the following complete Dirac equation, $(H - \varepsilon)\chi = 0$:

$$\left[ \begin{pmatrix} M-\varepsilon & \sigma_3(-\partial_r + V) \\ \sigma_3(\partial_r + V) & -M-\varepsilon \end{pmatrix} + \frac{\sigma_1}{r} \begin{pmatrix} 0 & -\partial_\theta + W \\ \partial_\theta + W & 0 \end{pmatrix} \right.$$

$$\left. + \frac{\sigma_2}{r\sin\theta} \begin{pmatrix} 0 & -\partial_\phi + Q \\ \partial_\phi + Q & 0 \end{pmatrix} \right] \begin{pmatrix} g_+ \\ g_- \end{pmatrix} = 0 \qquad (2.12)$$

where $g_\pm = \begin{pmatrix} g_\pm^+ \\ g_\pm^- \end{pmatrix}$. If we write these spinor components as $g_s^\pm(\vec{r}) = R_s^\pm(r)\Theta_s^\pm(\theta)\Phi_s^\pm(\phi)$, where $s$ is the + or − sign, then Eq (2.12) becomes completely separated in all three coordinates as follows

$$\left(\pm \frac{d}{d\phi} + Q\right)\Phi_\pm = i\sigma_3 \varepsilon_\phi \Phi_\pm, \tag{2.13a}$$

$$\left(\pm \frac{d}{d\theta} + W - \frac{\varepsilon_\phi}{\sin\theta}\right)\Theta_\pm = \sigma_1 \varepsilon_\theta \Theta_\mp, \tag{2.13b}$$

$$\begin{pmatrix} M + \frac{\varepsilon_\theta}{r} - \varepsilon & \sigma_3\left(-\frac{d}{dr} + V\right) \\ \sigma_3\left(\frac{d}{dr} + V\right) & -M + \frac{\varepsilon_\theta}{r} - \varepsilon \end{pmatrix} \begin{pmatrix} R_+ \\ R_- \end{pmatrix} = 0, \tag{2.13c}$$

where $\varepsilon_\phi$ and $\varepsilon_\theta$ are the separation constants which are real and dimensionless. In the following two sections we construct the solution space for this set of equations and derive the relativistic energy spectrum.

### III. SOLUTION OF THE ANGULAR DIRAC EQUATION

For the case where $Q(\phi) = 0$ Eq. (2.13a) gives the normalized solution $\Phi_\pm(\phi) = \frac{1}{\sqrt{2\pi}} e^{\pm i\sigma_3 \varepsilon_\phi \phi} \begin{pmatrix} 1 \\ 1 \end{pmatrix}$. That is,

$$\Phi_s^\pm(\phi) = \frac{1}{\sqrt{2\pi}} e^{\pm i s \varepsilon_\phi \phi}, \tag{3.1}$$

where again the sign $s$ = + or −. The requirement that $f_\pm(\phi) = f_\pm(\phi + 2\pi)$ results in a constraint on the real values of $\varepsilon_\phi$. Now, since $f_\pm(\vec{r}) = \Lambda(\hat{r})g_\pm(\vec{r})$, then this constraint dictates that $e^{\pm i 2\pi \varepsilon_\phi} = -1$. Hence, we should have

$$\varepsilon_\phi = \frac{m}{2}, \qquad m = \pm 1, \pm 3, \pm 5, \ldots \tag{3.2}$$

On the other hand, Eq. (2.13b) results in the following second order differential equation

$$\left[\frac{d^2}{d\theta^2} - W^2 \pm \frac{dW}{d\theta} + 2\varepsilon_\phi \frac{W}{\sin\theta} + \varepsilon_\phi \frac{\pm\cos\theta - \varepsilon_\phi}{\sin^2\theta} + \varepsilon_\theta^2\right]\Theta_\pm = 0. \tag{3.3}$$

This resembles the supersymmetric quantum mechanical equation with super-potentials $\mathcal{W}^2 \pm \mathcal{W}'$ and eigenvalue $\varepsilon_\theta^2$, where $\mathcal{W} = \frac{\varepsilon_\phi}{\sin\theta} - W$ [9]. The method of supersymmetric quantum mechanics could be used to obtain the solution of this equation. Nonetheless, we employ an alternative approach that utilizes the findings in Ref. 7. First, we change variables to the configuration space with coordinate $x \in [-1, +1]$, where $x = \cos\theta$. Writing $W(\theta) = \frac{U(x)}{\sin\theta}$ casts Eq. (3.3) into the following form

$$\left[(1-x^2)\frac{d^2}{dx^2} - x\frac{d}{dx} \mp \frac{dU}{dx} - \frac{(U-\varepsilon_\phi)(U-\varepsilon_\phi \pm x)}{1-x^2} + \varepsilon_\theta^2\right]\Theta_\pm = 0. \tag{3.4}$$

The solution of this equation (with $U \to -U$) has already been found in Sec. II of Ref. 7. That is with $U(x) = bx - a$, where $a$ and $b$ are real dimensionless parameters, we could write $\Theta_\pm(\theta)$ in terms of the Jacobi polynomials $P_n^{(\mu,\nu)}(x)$ [10] and as follows

$$\Theta_\pm(\theta) = A_n (1-x)^\alpha (1+x)^\beta P_n^{(\mu,\nu)}(x), \tag{3.5}$$

where $n = 0, 1, 2, \ldots$ and

$$A_n = \sqrt{\frac{2n+\mu+\nu+1}{2^{\mu+\nu+1}} \frac{\Gamma(n+1)\Gamma(n+\mu+\nu+1)}{\Gamma(n+\mu+1)\Gamma(n+\nu+1)}}, \qquad (3.6a)$$

$$\mu = \left|a - b + \tfrac{m\mp 1}{2}\right|, \quad \nu = \left|a + b + \tfrac{m\pm 1}{2}\right|, \qquad (3.6b)$$

$$\alpha = \tfrac{\mu+1/2}{2}, \quad \beta = \tfrac{\nu+1/2}{2}, \qquad (3.6c)$$

$$\varepsilon_\theta^2 = \left(n + \tfrac{\mu+\nu+1}{2}\right)^2 - b^2. \qquad (3.6d)$$

The top and bottom sign in these formulas goes with the corresponding subscript of $\Theta_\pm$. The right hand side of Eq. (3.6d) is positive for all real values of $a$ and $b$ and for all integers $m$. Now, for a given $\varepsilon_\theta$ the odd integer $m$ could, in principle, assume any value in the range $m = \pm 1, \pm 3, \pm 5, \ldots, \pm \hat{m}$, where $\hat{m}$ is the maximum positive odd integer satisfying

$$\hat{m} \leq -1 - 2|a| + 2\sqrt{\varepsilon_\theta^2 + b^2}. \qquad (3.7)$$

However, for $b \neq 0$ we should exclude from this range the following set of odd integers

$$m \notin \{-2(|b|+a) - 1 < m < 2(|b|-a) + 1\}. \qquad (3.8)$$

Moreover, for any odd integer $m$ in this permissible range the non-negative integer $n$ is determined from Eq. (3.6d) as

$$n = \sqrt{\varepsilon_\theta^2 + b^2} - \left|a + \tfrac{m}{2}\right| - \tfrac{1}{2} = 0, 1, 2, \ldots \qquad (3.9)$$

Collecting all of the above, we could rewrite the angular component $\Theta_\pm(\theta)$ in Eq. (3.5) as follows [7]

$$\Theta_\pm^s(\theta) = A_n \sqrt{\sin\theta}\, (1-x)^{\frac{1}{2}\left|a-b+\frac{m\mp 1}{2}\right|} (1+x)^{\frac{1}{2}\left|a+b+\frac{m\pm 1}{2}\right|} P_n^{(|a-b+\frac{m\mp 1}{2}|, |a+b+\frac{m\pm 1}{2}|)}(x). \qquad (3.10)$$

Again, all of these results are derivable from Sec. II of Ref. 7. In the following section we obtain the radial component of the spinor wavefunction by solving Eq. (2.13c) for the case where $V(r) = \lambda/r$ and $\lambda$ is a real coupling parameter. The energy spectrum for the bound states will also be obtained.

## IV. SOLUTION OF THE RADIAL DIRAC EQUATION AND THE ENERGY SPECTRUM

The radial Dirac equation (2.13c) with the potential $V(r) = \lambda/r$ could be rewritten in the following form

$$\begin{pmatrix} M + \frac{\varepsilon_\theta}{r} - \varepsilon & -\frac{d}{dr} + \frac{\lambda}{r} \\ \frac{d}{dr} + \frac{\lambda}{r} & -M + \frac{\varepsilon_\theta}{r} - \varepsilon \end{pmatrix} \begin{pmatrix} R_+ \\ \sigma_3 R_- \end{pmatrix} = 0. \qquad (4.1)$$

This matrix equation results in two coupled first order differential equations for the two radial spinor components. We make a unitary transformation of (4.1) such that by eliminating one component in favor of the other we obtain a second order differential equation that is Schrödinger-like (i.e., it contains no first order derivatives). This is accomplished by applying the transformation $e^{\frac{i}{2}\eta\sigma_2}$ on the radial Dirac equation (4.1), where $\eta$ is a real constant parameter. The Schrödinger-like requirement dictates that the parameter $\eta$ should satisfy the constraint that $\sin(\eta) = \pm \varepsilon_\theta/\lambda$, where $-\tfrac{\pi}{2} \leq \eta \leq +\tfrac{\pi}{2}$ depending on the signs of $\lambda$ and $\varepsilon_\theta$. Consequently, Eq. (4.1) gets transformed into the following

$$\begin{pmatrix} \frac{M\gamma}{\lambda} + (1\pm 1)\frac{\varepsilon_\theta}{r} - \varepsilon & \mp\frac{M\varepsilon_\theta}{\lambda} + \frac{\gamma}{r} - \frac{d}{dr} \\ \mp\frac{M\varepsilon_\theta}{\lambda} + \frac{\gamma}{r} + \frac{d}{dr} & -\frac{M\gamma}{\lambda} + (1\mp 1)\frac{\varepsilon_\theta}{r} - \varepsilon \end{pmatrix} \begin{pmatrix} \tilde{R}_+(r) \\ \tilde{R}_-(r) \end{pmatrix} = 0, \qquad (4.2)$$

where $\gamma = \frac{\lambda}{|\lambda|}\sqrt{\lambda^2 - \varepsilon_\theta^2}$ and

$$\begin{pmatrix} \tilde{R}_+ \\ \tilde{R}_- \end{pmatrix} = e^{\frac{i}{2}\eta\sigma_2} \begin{pmatrix} R_+ \\ \sigma_3 R_- \end{pmatrix} = \begin{pmatrix} \cos\frac{\eta}{2} & \sin\frac{\eta}{2} \\ -\sin\frac{\eta}{2} & \cos\frac{\eta}{2} \end{pmatrix} \begin{pmatrix} R_+ \\ \sigma_3 R_- \end{pmatrix}. \qquad (4.3)$$

The expression for $\gamma$ above shows that real solutions are obtained only if the integers $n$ and $m$ assume values such that $|\varepsilon_\theta| \leq |\lambda|$. Equation (4.2) gives one radial spinor component in terms of the other as follows

$$\tilde{R}_\mp = \frac{1}{(\gamma/\lambda)M \pm \varepsilon}\left(-M\frac{\varepsilon_\theta}{\lambda} \pm \frac{\gamma}{r} + \frac{d}{dr}\right)\tilde{R}_\pm, \qquad (4.4)$$

where $\varepsilon \neq \mp(\gamma/\lambda)M$. This equation is referred to as the "kinetic balance" relation. Since $\frac{\gamma}{\lambda} > 0$ then $\varepsilon = +\frac{\gamma}{\lambda}M$ ($\varepsilon = -\frac{\gamma}{\lambda}M$) is an element of the positive (negative) energy spectrum. Therefore, this relation with the top (bottom) sign is not valid for the negative (positive) energy solutions. Our choice is to construct the positive energy solutions. The negative energy solutions, on the other hand, are obtained from these positive energy solutions simply by the map

$$\varepsilon \to -\varepsilon, \; \lambda \to -\lambda, \; \varepsilon_\theta \to -\varepsilon_\theta, \; \tilde{R}_\pm \leftrightarrow \tilde{R}_\mp. \qquad (4.5)$$

One can easily verify that this map transforms the kinetic balance relation (4.4) and Eq. (4.2) with the top sign into the same but with the bottom sign. Now, to obtain the positive energy solutions, where $\sin(\eta) = +\varepsilon_\theta/\lambda$, we use Eq. (4.4) to eliminate the lower radial component in favor of the upper in Eq. (4.2) with the top sign. The resulting Schrödinger-like second order differential equation is[+]

$$\left[-\frac{d^2}{dr^2} + \frac{\gamma(\gamma+1)}{r^2} + 2\varepsilon\frac{\varepsilon_\theta}{r} - (\varepsilon^2 - M^2)\right]\tilde{R}_+(r) = 0, \qquad (4.6)$$

which shows that $\gamma$ plays the role of the angular momentum quantum number, albeit not an integer. We propose the following solution

$$\tilde{R}_+(r) = C_k(\xi r)^\rho e^{-\xi r/2} L_k^\tau(\xi r), \qquad (4.7)$$

where $L_k^\tau(z)$ is the Laguerre polynomial of order $k$ ($k = 0, 1, 2, ...$) and $C_k$ is the normalization constant. The real parameter $\xi$ is positive with inverse length dimension, whereas, $\rho$ and $\tau$ are dimensionless such that $\rho > 0$ and $\tau > -1$. These constraints on the parameters make the proposed solution square integrable and compatible with the boundary conditions. Substituting (4.7) into Eq. (4.6) and using the differential equation of the Laguerre polynomials [10] we obtain (depending on the sign of $\lambda$) the following

---

[+] The corresponding negative energy equation is obtained by choosing $\sin(\eta) = -\varepsilon_\theta/\lambda$ and using the kinetic balance relation (4.4) to eliminate the upper radial component in Eq. (4.2) with the bottom sign giving

$$\left[-\frac{d^2}{dr^2} + \frac{\gamma(\gamma-1)}{r^2} + 2\varepsilon\frac{\varepsilon_\theta}{r} - (\varepsilon^2 - M^2)\right]\tilde{R}_-(r) = 0.$$

This shows that $\gamma - 1$ plays the role of the angular momentum quantum number. Additionally, this equation could have also been obtained by the action of the map (4.5) on Eq. (4.6).

positive energy spectrum and parameter values that produce a diagonal representation in the solutions space

$$\rho_\pm = \begin{cases} \gamma+1 &, \lambda > 0 \\ -\gamma &, \lambda < 0 \end{cases}, \quad \varepsilon_{knm}^\pm = M\left[1+\left(\frac{\varepsilon_\theta}{k+\rho_\pm}\right)^2\right]^{-\frac{1}{2}}, \tag{4.8}$$

$$\tau_\pm = 2\rho_\pm - 1, \quad \xi_\pm = -2\varepsilon_{knm}^\pm \varepsilon_\theta / (k+\rho_\pm). \tag{4.9}$$

The top and bottom sign corresponds to $\pm\lambda > 0$. The dependence on the indices $n$ and $m$ comes from $\varepsilon_\theta$ whose magnitude could be written, using Eq. (3.6d) and the constraint (3.8), as

$$|\varepsilon_\theta| = \sqrt{\left(n+\left|\frac{a+m}{2}\right|+\frac{1}{2}\right)^2 - b^2}. \tag{4.10}$$

It is obvious that $\varepsilon_{k+1,n,m}^-\big|_\gamma = \varepsilon_{k,n,m}^+\big|_{-\gamma}$. Thus, the energy spectrum is two-fold degenerate and could be written collectively as follows:

$$\varepsilon_{knm} = M\left[1+\left(\frac{\varepsilon_\theta}{k+1+\sqrt{\lambda^2-\varepsilon_\theta^2}}\right)^2\right]^{-\frac{1}{2}}. \tag{4.11}$$

The only non-degenerate positive energy state is the one associated with the highest energy (upper bound of the spectrum) given by Eq. (4.8) for $k = 0$ and $\lambda < 0$ as $\varepsilon_{0nm}^-$, which is equal to $M\gamma/\lambda$. On the other hand, there exists another non-degenerate but negative energy state associated with the lowest energy (lower bound of the spectrum) where $\varepsilon = -M\gamma/\lambda$. It is obtained from the positive energy solutions by the map (4.5). Now, we can write the normalized upper component of the radial spinor wave function as

$$\tilde{R}_+(r) = \begin{cases} \sqrt{\frac{\xi_{knm}\Gamma(k+1)}{\Gamma(k+2\gamma+2)}}(\xi_{knm}r)^{\gamma+1} e^{-\xi_{knm}r/2} L_k^{2\gamma+1}(\xi_{knm}r) &, \lambda > 0 \\ \sqrt{\frac{\xi_{knm}\Gamma(k+2)}{\Gamma(k-2\gamma+1)}}(\xi_{knm}r)^{-\gamma} e^{-\xi_{knm}r/2} L_{k+1}^{-2\gamma-1}(\xi_{knm}r) &, \lambda < 0 \end{cases} \tag{4.12}$$

where $k = 0,1,2,..$, $\xi_{knm} = -2\varepsilon_{knm}\varepsilon_\theta/\left(k+1+\sqrt{\lambda^2-\varepsilon_\theta^2}\right)$ and $\varepsilon_\theta < 0$. The non-degenerate state associated with the highest energy, $\varepsilon = M\gamma/\lambda$, has the following upper radial spinor component

$$\tilde{R}_+(r) = \begin{cases} 0 &, \lambda > 0 \\ \sqrt{\frac{2M|\varepsilon_\theta/\lambda|}{\Gamma(-2\gamma)}} \left(2M|\varepsilon_\theta/\lambda|r\right)^{-\gamma} e^{-M|\varepsilon_\theta/\lambda|r} &, \lambda < 0 \end{cases} \tag{4.13}$$

The lower radial component $\tilde{R}_-$ of the positive energy solution is obtained from the upper in (4.12) and (4.13) using the kinetic balance relation (4.4) with the top sign. On the other hand, the radial components of the spinor wavefunction associated with the negative energy solutions are obtained from the positive energy solutions above using the map (4.5). Finally, we obtain the total positive-energy four-component spinor wavefunction as follows:

$$\psi(\vec{r}) = \frac{1}{r\sqrt{\sin\theta}}\begin{pmatrix} if_+ \\ f_- \end{pmatrix} = \frac{1}{r\sqrt{\sin\theta}}\begin{pmatrix} i\Lambda g_+ \\ \Lambda g_- \end{pmatrix} = \frac{1}{r\sqrt{2\pi\sin\theta}}\begin{pmatrix} i\Lambda\begin{pmatrix} e^{\frac{i}{2}m\phi} \\ e^{-\frac{i}{2}m\phi} \end{pmatrix}\Theta_+(\omega_+\tilde{R}_+ - q\omega_-\tilde{R}_-) \\ \Lambda\begin{pmatrix} e^{-\frac{i}{2}m\phi} \\ -e^{\frac{i}{2}m\phi} \end{pmatrix}\Theta_-(\omega_+\tilde{R}_- + q\omega_-\tilde{R}_+) \end{pmatrix}, \tag{4.14}$$

where $\omega_\pm = \sqrt{\frac{1}{2}(1\pm\gamma/\lambda)}$, $q = \text{sign}(\lambda\varepsilon_\theta) = \lambda\varepsilon_\theta/|\lambda\varepsilon_\theta|$.

## ACKNOWLEDGMENTS

I am grateful to M. S. Abdelmonem and H. Bahlouli for their help in literature search.